\documentclass[aps,prl,reprint,superscriptaddress,floatfix]{revtex4-1}
\usepackage{amssymb,bm}
\usepackage{amsmath,eucal,ulem}
\usepackage[usenames,dvipsnames]{color}
\usepackage{graphicx}
\begin{document}

\title{Anomalous thermodynamics at the micro-scale} 
\author{Antonio Celani}
\affiliation{Physics of Biological Systems, Institut Pasteur and CNRS UMR 3525, 28 rue du docteur Roux, 75015 Paris, France} 
\author{Stefano Bo}
\affiliation{Cancer Cell Biophysics, Institute for Cancer Research and Treatment, Str.~Prov.~142~km~3.95, 10060 Candiolo,
Torino, Italy}
\affiliation{INFN, via P. Giuria 1, 10125 Torino, Italy} 
\affiliation{Dept. Computational Biology, AlbaNova University Centre, KTH - Royal Institute of Technology, SE-106 91 Stockholm, Sweden}
\affiliation{NORDITA- Roslagstullsbacken 23, SE-106 91 Stockholm, Sweden}
\author{Ralf Eichhorn}
\affiliation{NORDITA- Roslagstullsbacken 23, SE-106 91 Stockholm, Sweden}
\author{Erik Aurell}
\affiliation{Dept. Computational Biology, AlbaNova University Centre, KTH - Royal Institute of Technology, SE-106 91 Stockholm, Sweden}
\affiliation{ACCESS Linnaeus Centre, KTH - Royal Institute of Technology, SE-100 44 Stockholm, Sweden}
\affiliation{Dept Information and Computer Science, Aalto University, PO Box 15400, FI-00076 Aalto, Finland}
\begin{abstract}
Particle motion at the micro-scale is an incessant tug-of-war between thermal fluctuations and applied forces on one side, and the strong resistance exerted by fluid viscosity on the other. Friction is so strong that completely neglecting inertia -- the overdamped approximation -- gives an excellent effective description of the actual particle mechanics. In sharp contrast with this result, here we show that the overdamped approximation dramatically fails when thermodynamic quantities such as the entropy production in the environment are considered, in presence of temperature gradients. In the limit of vanishingly small, yet finite inertia, we find that the entropy production is dominated by a contribution that is anomalous, i.e. has no counterpart in the overdamped approximation. This phenomenon, that we call 
entropic anomaly, is due to a symmetry-breaking that occurs when moving to the small, finite inertia limit.  
{
Anomalous entropy production is traced back to futile phase-space cyclic trajectories displaying a fast downgradient sweep followed by a slow upgradient return to the original position.
}
\end{abstract}
\date{\today}
\maketitle

Life at the micro-scale flows under one law: the fluid gives and the fluid takes away. This arbitrary tyrant lavishly bestows energy to suspended particles through molecular collisions while incessantly draining it from them through friction. The result is the erratic movment of microscopic particles that goes under the name of Brownian motion.

The theory of Brownian motion was developed by Einstein, Smoluchowski and Langevin a little over a century ago~\cite{Duplantier}. A central result of this theory is the {\it overdamped} approximation, which says that inertia can be ignored if mass
is small, or friction is large. 
The motion of a Brownian particle, which obeys Newtonian mechanics and is
driven by collisions and external forces, is thus reduced to a first-order diffusion equation. 
The overdamped approximation successfully describes
the mechanics on the microscale and is very widely used~\cite{P77,Dusenbery}.

With the advent of micromanipulation it has become possible to measure and control the positions of individual
Brownian particles and other small systems~\cite{Ritort06,LKM10,BB11,BAP12}. Thermodynamic concepts such as heat, work and entropy production have hence taken a meaning for single systems~\cite{S05,Sekimoto}. These developments are the
foundation of the new emerging field of stochastic thermodynamics, where the most striking results obtained to date are fluctuation relations, recently reviewed in, e.g., Refs.~\cite{Jarzynski08,Esposito09,Jarzynski11}. Here we show 
that in this setting the overdamped approximation fails, as soon as the temperature field varies in space.

Indeed, although the overdamped approximation correctly yields the trajectories of the 
Brownian particles in space,
it incorrectly estimates the entropy production. This failure is traced back to seemingly innocent corrections that, while having a negligible impact on trajectories, eventually dominate the entropy production in the long run, even in the limit when inertial
effects go to zero. 
As a result, fluctuation relations themselves take a very nontrivial form in the limit of small yet finite inertia.

We dub this phenomenon {\it entropic anomaly}, in analogy with similar anomalies encountered in physics.
The best known example in classical physics is the viscous dissipative anomaly. The energy dissipation in a fluid flow remains finite even in the limit of arbitrarily small viscosity,
whereas it vanishes when viscosity is exactly zero \cite{Frisch}.
The viscous anomaly reflects the loss of time-reversal symmetry of fluid dynamics when going from the inviscid case to the viscous one. Quantum anomalies arise when a current conserved at the classical level (Planck's constant set to zero) is not conserved anymore at the quantum level (see, e.g., Ref.~\cite{FS04} for a general discussion and Ref.~\cite{CEF01} for a simple example from molecular physics). 
We show here that the entropic anomaly is associated to the breaking of a symmetry of the zero-inertia, overdamped dynamics, when small finite inertia is considered. The symmetry is the joint reversal of time and particle velocity.

To illustrate our result, let us focus on a minimal thought experiment. 
A microscopic particle is suspended in a fluid inside a vessel with reflecting walls. The two opposite sides of the vessel are in contact with heat reservoirs at 
different temperatures. 
The fluid is motionless and displays by a static, smooth temperature profile $T({\bm x})$ (e.g. a linear one). 
For the sake of simplicity, no external force is applied to the particle 
and the friction coefficient is taken independent of particle position. 
We emphasize that the
assumptions on temperature profiles, friction coefficients and absence of external forces
are not restrictive, and refer the discussion below and the Supplementary Material for a general discussion.

The motion of the suspended particle is then governed by the Langevin-Kramers equations for position ${\bm X}_t$ and velocity ${\bm V}_t$ \cite{Dimensions} 
\begin{equation}\label{eq:LK}
\begin{array}{lll}
\dot{\bm X}_t &=& {\bm V}_t \\
\dot{\bm V}_t &=& -\gamma {\bm V}_t + \sqrt{2 T({\bm X}_t) \gamma}\, {\bm \eta}_t
\end{array}
\end{equation}
where ${\bm \eta}_t$ is a Gaussian, zero-mean white noise, i.e. $\langle \eta^i_t\eta^j_{t'} \rangle =\delta^{ij}\delta(t-t')$.

The heat released by the particle to the fluid along a trajectory from time $t'$ to time $t$ reads
$Q =  -|{\bm V}_t|^2/2 + |{\bm V}_{t'}|^2/2 $.
In the case at hand, since there is no force, no work is done on the particle and no potential energy is stored~\cite{Note}.
The entropy of the particle is defined as the state function
\begin{equation}\label{eq:Sp}
S_{p}({\bm x},{\bm v},t) = -\ln p ({\bm x},{\bm v},t)
\end{equation}
where $p$ is the probability density of the particle position and velocity at time $t$ ~\cite{S05}. In other words, $p$ is the solution of the Fokker-Planck equation associated to Eq.~\eqref{eq:LK}. The entropy produced by the particle in the environment (the fluid) along a path is the integral of the released heat divided 
by temperature
\begin{equation}\label{eq:Senv}
S_{\mathit{env}} = - \int_{t'}^t  \frac{1}{T({\bm X}_\tau)} {\bm V}_\tau \circ \dot{\bm V}_\tau \,
d\tau  
\end{equation}
where $\circ$  denotes the Stratonovich product (the integral is regularized according to the midpoint rule).
The total entropy produced along the process is therefore 
\begin{equation}\label{eq:Stot}
S_{tot} =S_p({\bm X}_t,t)-S_p({\bm X}_{t'},t') + S_{env}
\end{equation}
and  is positive on average as prescribed by the second law of thermodynamics \cite{CG08,Note2}.

We now turn our attention to the overdamped dynamics. It is possible to prove by means of asymptotic techniques (see \cite{Sekimoto,B97,MS00} and the Supplemental Material)  that the the spatial trajectory ${\bm X}_t$ in \eqref{eq:LK} tends -- in the limit of small inertia and in the probabilistic sense -- to the solution of 
\begin{equation}\label{eq:OD}
\begin{array}{lll}
\dot{\bm X}_t & = & \sqrt{2 T({\bm X}_t) / \gamma } \; {\bm \eta}_t
\end{array}
\end{equation}
The correct interpretation of the equation above is that the product on the right-hand-side has to be taken with the It\=o, nonanticipative convention. 
The velocity follows the local Maxwell-Boltzmann
distribution $w({\bm v}|{\bm x})=(2\pi T({\bm x}))^{-3/2} \exp[-|{\bm v}|^2/(2T({\bm x}))]$.

The Fokker-Planck equation associated to Eq.~\eqref{eq:OD} can be interpreted as
the mass-conservation equation for a dilute colloidal suspension of noninteracting particles.
The flux of particles is  ${\bm J} = -T \gamma^{-1} {\bm \nabla} \rho - \gamma^{-1}\rho
{\bm \nabla} T$ and features the contributions of the osmotic force $-T {\bm \nabla} \ln \rho$, and of the thermophoretic force $- {\bm \nabla} T$. 
For times larger than $L^2\gamma/T$, where $L$ is the size of the vessel that encloses the fluid, the probability density of particle position reaches  the equilibrium $\rho_{eq} \propto T^{-1}$ with a zero-flux balance of  osmotic and thermophoretic forces.

Stochastic thermodynamics can be formulated for the overdamped dynamics \eqref{eq:OD}.
The particle entropy is
\begin{equation}\label{eq:ODSp}
S^{\mathit{(over)}}_p = -\ln \rho({\bm x},t) 
\end{equation}
where $\rho$ solves the Fokker-Planck equation associated to Eq.~\eqref{eq:OD}. When the particle moves from a region where the probability density $\rho$ is high 
to one where it is low, its entropy increases.
The entropy produced in the environment is the work done by the thermophoretic force on the particle
along a trajectory, divided by temperature
\begin{equation}\label{eq:ODSenv}
S^{\mathit{(over)}}_{env} = -\int_{t'}^t \frac{{\bm \nabla} T ({\bm X}_\tau)}{T({\bm X}_\tau)} \circ \dot{\bm X}_\tau d\tau = \ln \left( \frac{T({\bm X}_{t'})}{T({\bm X}_{t})} \right) 
\end{equation}
and it is positive when the particle moves from a hot region 
into a cold one.
When the system
reaches equilibrium the total entropy production 
\begin{equation}\label{eq:ODStot}
S_{tot}^{(over)}= S^{\mathit{(over)}}_p({\bm X}_t,t) - S^{\mathit{(over)}}_p({\bm X}_{t'},t') +S^{\mathit{(over)}}_{env}
\end{equation}
 tends to zero along any trajectory since these two contributions compensate exactly.
The average rate of entropy production is  
\begin{equation}\label{eq:ODSave}
\frac{d}{dt} \left\langle S_{\mathit{tot}}^{\mathit{(over)}} \right\rangle = 
\int \frac{\left|{\bm \nabla} (\rho T)\right|^2}{\gamma \rho T} d{\bm x} \ge 0
\end{equation}
where the integral is over the volume of the vessel. It vanishes in equilibrium since $\rho T$ is then a constant.
This result coincides with the expression obtained by macroscopic non-equilibrium thermodynamics
for the entropy production by a dilute particle suspension \cite{GM}.

We now move to the
description of the results of the present work, where the limit of small {\it yet finite} inertia
is considered. The derivations follow standard asymptotic expansion methods and are detailed in the Supplementary Material. The only assumption is that the length-scale of variation of the temperature be larger than
the typical distance travelled during the friction time $\gamma^{-1}$, i.e. the gradients
must not be exceedingly large $|{\bm \nabla} T| \ll  T \gamma / v \sim \gamma T^{1/2}$
\cite{Note-BL-ratchet}.

The average rate of entropy production of the Langevin-Kramers process \eqref{eq:LK}, in the limit of vanishing inertia, is
\begin{equation}\label{eq:Save}
\frac{d}{dt}\left\langle S_{\mathit{tot}} \right\rangle =  \frac{d}{dt}\left\langle S^{\mathit{(over)}}_{\mathit{tot}} \right\rangle + \frac{5}{6} \left\langle  \frac{T}{\gamma} \left(\frac{{\bm \nabla} T}{T} \right)^2 \right\rangle
\end{equation}
This {\it exact asymptotic} expression has to be contrasted with the overdamped {\it approximation}
\eqref{eq:ODSave}. It differs by an additional, positive contribution, which
actually controls the asymptotic rate of entropy production since $d\langle S_{\mathit{tot}} ^{\mathit{(over)}}\rangle/dt$ vanishes as equilibrium is approached. 
This is the most conspicuous effect of the entropic anomaly.

It is indeed possible to isolate the source of the anomaly from the total entropy production
\eqref{eq:Stot}, $S_{\mathit{tot}}=S_{\mathit{reg}}+S_{\mathit{anom}}$, where
\begin{equation}
S_{\mathit{anom}} = \int_{t'}^t \frac{(5 T({\bm X}_\tau) -\left|{\bm V}_\tau\right|^2)}{2T({\bm X}_\tau)
^2} {\bm V}_\tau \cdot {\bm \nabla} T({\bm X}_\tau) \,d\tau 
\end{equation}
and to show that it arises from the $S_{\mathit{env}}$ contribution. The remainder $S_{\mathit{reg}}$ has a regular limit and tends to the overdamped entropy \eqref{eq:ODStot}. 
In the limit of small inertia 
the anomalous entropy obeys the identity
\begin{equation}\label{eq:Jarz-anom}
\left\langle \exp\left(-S_{\mathit{anom}}\right) \right\rangle =1\;
\end{equation}
and its average rate of production is
\begin{equation}\label{eq:Sanomave}
\frac{d}{dt} \left\langle S_{\mathit{anom}} \right\rangle = 
\frac{5}{6}\int \rho \frac{ ({\bm \nabla} T)^2 }{\gamma T} d{\bm x}
\end{equation}
that gives the rightmost term in \eqref{eq:Save}. 
It is worth pointing out that the anomalous contribution cannot be eliminated by a suitable redefinition of the overdamped entropy.  Indeed, it does not exist any sequential functional of the trajectories \eqref{eq:OD} that gives the correct limiting statistics of  $S_{\mathit{anom}}$
(see Supplemental Material).
  
\begin{figure}[!t]
\includegraphics[width=\columnwidth, trim= 100 100 100 100]{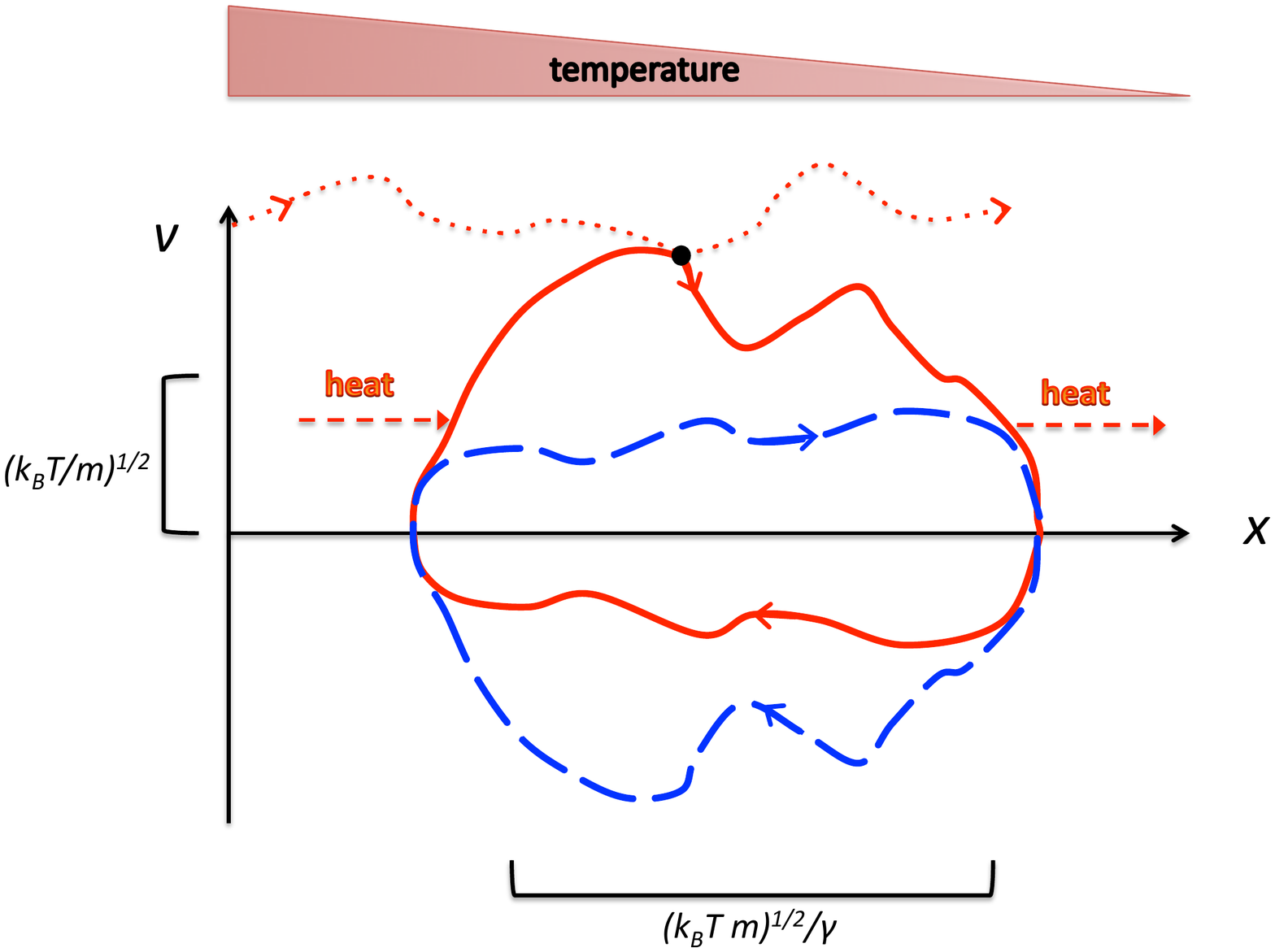}
\label{fig:1}
\caption{
{
Anomalous entropy is produced during futile cycles of fast downgradient sweeping and slow upgradient climbing.  Along a microscopic trajectory that starts and ends at the same temperature (full red line), the overdamped contribution to entropy vanishes according to eq.~\eqref{eq:ODSenv}. However, anomalous entropy is produced because of heat absorption at higher temperatures, followed by heat release at smaller ones. The reversed cycle (blue dashed line) with negative anomalous entropy production is much less likely to occur in agreement with eq.~\eqref{eq:condit}.
}
}
\end{figure}
 
What is the interpretation of the anomaly~?
Entropy is a measure of irreversibility. It is possible to show in very general terms that
the entropy produced in the environment along a trajectory 
quantifies the probability of the reversed path relative to the forward one (see \cite{CG08}
for a precise statement for diffusion processes).
In the Langevin-Kramers case, the reversal consists in the inversion of the arrow of time and 
of the direction of velocity (see Figure~1) and one has
\begin{equation*}
\frac{P({\bm X}_{[t,t']},-{\bm V}_{[t,t']})}{P({\bm X}_{[t',t]},{\bm V}_{[t',t]})}= \exp\left(-S_{\mathit{env}} \right)
\end{equation*}
Similarly, for the overdamped dynamics,
\begin{equation*}
\frac{P^{\mathit(over)}({\bm X}_{[t,t']})}{P^{\mathit(over)}({\bm X}_{[t',t]})}= \exp\left(-S^{\mathit(over)}_{\mathit{env}} \right)
\end{equation*}
From the ratio of these two identities it follows that, in the limit of vanishing inertia, the anomalous entropy gives the relative weight of {\it conditional} probabilities 
\begin{equation}\label{eq:condit}
\frac{P(-{\bm V}_{[t,t']}|{\bm X}_{[t,t']})}{P({\bm V}_{[t',t]}|{\bm X}_{[t',t]})}= \exp\left(-S_{\mathit{anom}} \right)
\end{equation}
This result shows that the anomaly emerges from the breaking of time and velocity reversal symmetry at a given spatial position. 

The anomaly can be traced back to the deviation of the velocity distribution
from the Maxwell-Boltzmann one, $w({\bm v}|{\bm x})$. 
At equilibrium, the velocity statistics is indeed
given by 
\begin{equation}\label{eq:maxwdev}
p({\bm v} | {\bm x}) = w({\bm v}|{\bm x}) \left( 1 + \frac{(5 T({\bm x})-|{\bm v}|^2)} {6 \gamma T({\bm x})^2}  {\bm v} \cdot {\bm \nabla} T({\bm x}) \right)
\end{equation}
except for terms of order $\gamma^{-2}$ or higher.
This deviation, albeit small, plays a crucial role as it breaks the velocity reversal symmetry
along the temperature gradient direction, and is therefore responsible
for the anomalous entropy production.

{ Further insight on the interpretation of the anomalous entropy can be gained
by the following argument (see Figure~\ref{fig:1}).
At the microscopic level (scales of order of  $T^{1/2}/\gamma$, times of order 
$1/\gamma$) the particle trajectory often ``goes round in circles''.
Indeed, as seen from eq.~\eqref{eq:maxwdev}, 
the trajectories with higher probability (positive sign of the correction term) are characterized by a downgradient sweep with speeds larger than the thermal velocity, followed by a slow upgradient motion (red trajectory in Figure~\ref{fig:1}). Note that the correction does not contribute to the spatial flux of particles as can be directly checked by multiplying by ${\bm v}$ and integrating over velocity. 
It does not contribute to the mean kinetic heat exchange either, since it is 
odd in velocity coordinates.
During this futile cycling, however, heat is absorbed at
higher temperatures and released at smaller ones,  thereby producing entropy.
In each cycle of duration $\sim \gamma^{-1}$, 
the amount of entropy produced is approximately given by the heat exchanged 
$\sim T$, times the difference of inverse temperatures at the extremes of the trajectory, 
$\sim T^{-2} |\nabla T| (T^{1/2}/\gamma)$. Reversed cycles (such as the one depicted in  blue in Figure~\ref{fig:1}) give a similar contribution with opposite sign. However, forward cycles are more 
likely with probability $\sim T^{-1/2} |\nabla T|/\gamma$ as follows from eq.~\eqref{eq:maxwdev}.
This results in an overall average entropy production rate $\sim  \gamma^{-1} T^{-1} |\nabla T|^2$ that we recognize as the anomalous term in eq.~\eqref{eq:Save}. 
}

The present findings have a much broader range of applicability
than the simple example discussed here (see Supplementary material).
The addition of potential and nonconservative forces is straightforward
since the related heat, work and entropy contributions
all have regular limits and are thus correctly described by the overdamped dynamics. 
A position-dependent friction, which may account for the temperature dependence of viscosity or hydrodynamic effects due to the presence of material boundaries, can also be included
with minor changes. 
Finally, it is also possible to consider time-dependent temperature, friction and forces, provided the variation is not faster than the timescale of overdamped motion. 
All these modifications do not alter our results. 
 The interpretation is left unchanged as well. 
 
In conclusion, we have shown that thermodynamics at the micro-scale is intrinsically anomalous in the limit of vanishingly small inertia. The anomaly arises from the breaking of time-reversal symmetry that emerges when the particle is subject to a temperature gradient. 
{
How may the entropic anomaly impact observable phenomena~?
The most conspicuous effect should be on the efficiency of thermal stochastic
engines in nonuniform temperature environments.
Indeed, the intrinsic irreversibility arising from the anomalous
contribution -- which does not vanish even for quasistatic transformations --
should irremediably hamper the ability of the engine of converting absorbed heat into work.
Recent advances in experimental techniques of particle confinement, tracking, and heating might pave the way toward an experimental measurement of this effect \cite{BB11,BAP12,LKM10}.
}


\begin{thebibliography}{99}
\bibitem{Dimensions}
The general form of the Langevin-Kramers equation is
$\dot{\bm X}_t = {\bm V}_t$, $m\dot{\bm V}_t  = {\bm f} -\gamma {\bm v} + \sqrt{kT \gamma} {\bm \eta}_t$, where $m$ is the particle mass, ${\bm f}$ the applied force, $k$ the Boltzmann constant. Equation~\protect\eqref{eq:LK} follows by reduction to non-dimensional form through the rescalings  ${\bm X} \to {\bm X}/L$, ${\bm V} \to {\bm V}/\sqrt{kT_0/m}$, ${\bm f} \to {\bm f}/(kT_0/L)$, $t \to t/(L/\sqrt{kT_0/m})$, $\gamma \to \gamma L/\sqrt{kT_0m}$, $T \to T/T_0$, where $L$ is the characteristic length-scale of the problem
and $T_0$ is  a reference temperature. The overdamped approximation is obtained by setting the inverse of non-dimensional friction to zero, i.e. $\sqrt{kT_0m}/(\gamma L)=0$.
\bibitem{Note}
Note that the heat released to the fluid does not alter the temperature profile. This is because the thermal diffusivity of water in standard conditions is six orders of magnitude 
larger than diffusion coefficient of typical Brownian particles. The fluid recovers the steady state before the particle has travelled any appreciable distance.
\end{thebibliography}
\end{document}